\documentclass[prb,twocolumn,showpacs,amssymb,amsmath]{revtex4}

\usepackage{graphicx}
\usepackage{bm}
\def\be{\begin{equation}}
\def\ee{\end{equation}}
\def\bea{\begin{eqnarray}}
\def\eea{\end{eqnarray}}
\newcommand{\matr}[4]{{\left(\begin{array}{cc} #1&#2\\#3&#4\\\end{array}\right)}}
\newcommand{\vect}[2]{{\left(\begin{array}{c} #1\\#2\\\end{array}\right)}}
\renewcommand{\vec}{\mathbf}

\renewcommand{\vr}{\vec{r}}

\newcommand{\vsigma}{\mbox{\boldmath $\sigma$}}

\newcommand{\vA}{\vec{A}}
\newcommand{\vnabla}{\mbox{\boldmath $\nabla$}}

\newcommand{\vp}{\vec{p}}

\newcommand{\eps}{\varepsilon}
\newcommand{\vF}{v_{\mathrm{F}}}

\begin{document}
\title{Density of states as a probe of electrostatic confinement in graphene}

\author{Martin Schneider}
\author{Piet W. Brouwer}
\affiliation{
Dahlem Center for Complex Quantum Systems and Institut f\"ur Theoretische Physik,
Freie Universit\"at Berlin, Arnimallee 14, 14195 Berlin, Germany
}
\date{\today}

\begin{abstract}
We theoretically analyze the possibility to confine electrons in single-layer graphene with the help of metallic gates, via the evaluation of the density of states of such a gate-defined quantum dot in the presence of a ring-shaped metallic contact. The possibility to electrostatically confine electrons in a gate-defined ``quantum dot'' with finite-carrier density, surrounded by an undoped graphene sheet, strongly depends on the integrability of the electron dynamics in the quantum dot. With the present calculations we can quantitatively compare confinement in dots with integrable and chaotic dynamics, and verify the prediction that the Berry phase associated with the pseudospin leads to partial confinement in situations where no confinement is expected according to the arguments relying on the classical dynamics only.
\end{abstract}

\maketitle

\section{Introduction}

The controlled fabrication of high-quality nanostructures as well as its strictly two-dimensional nature have established graphene as an outstanding candidate for future nanoelectronic devices.\cite{novoselov2004,castroneto2009,geim2009,beenakker2008,peres2010,dassarma2011} At the same time, because of its gapless quasirelativistic dispersion, graphene lacks the possibility to create a depletion region by means of electrostatic gating. It is for this reason, that experimental activity aimed at confining electrons in graphene nanostructures focuses on quantum dots realized in etched structures.\cite{bunch2005,ponomarenko2008,stampfer2008,schnez2009,guttinger2012,jacobsen2012} The difficulty to confine electrons using electrostatic means is closely related to the phenomenon of Klein tunneling:\cite{klein1929,cheianov2006,katsnelson2006} Electrons that approach a potential barrier at perpendicular incidence will always be transmitted with unit probability, irrespective of the height and type of the barrier.

\begin{figure}[t]
\includegraphics[width=2.5in]{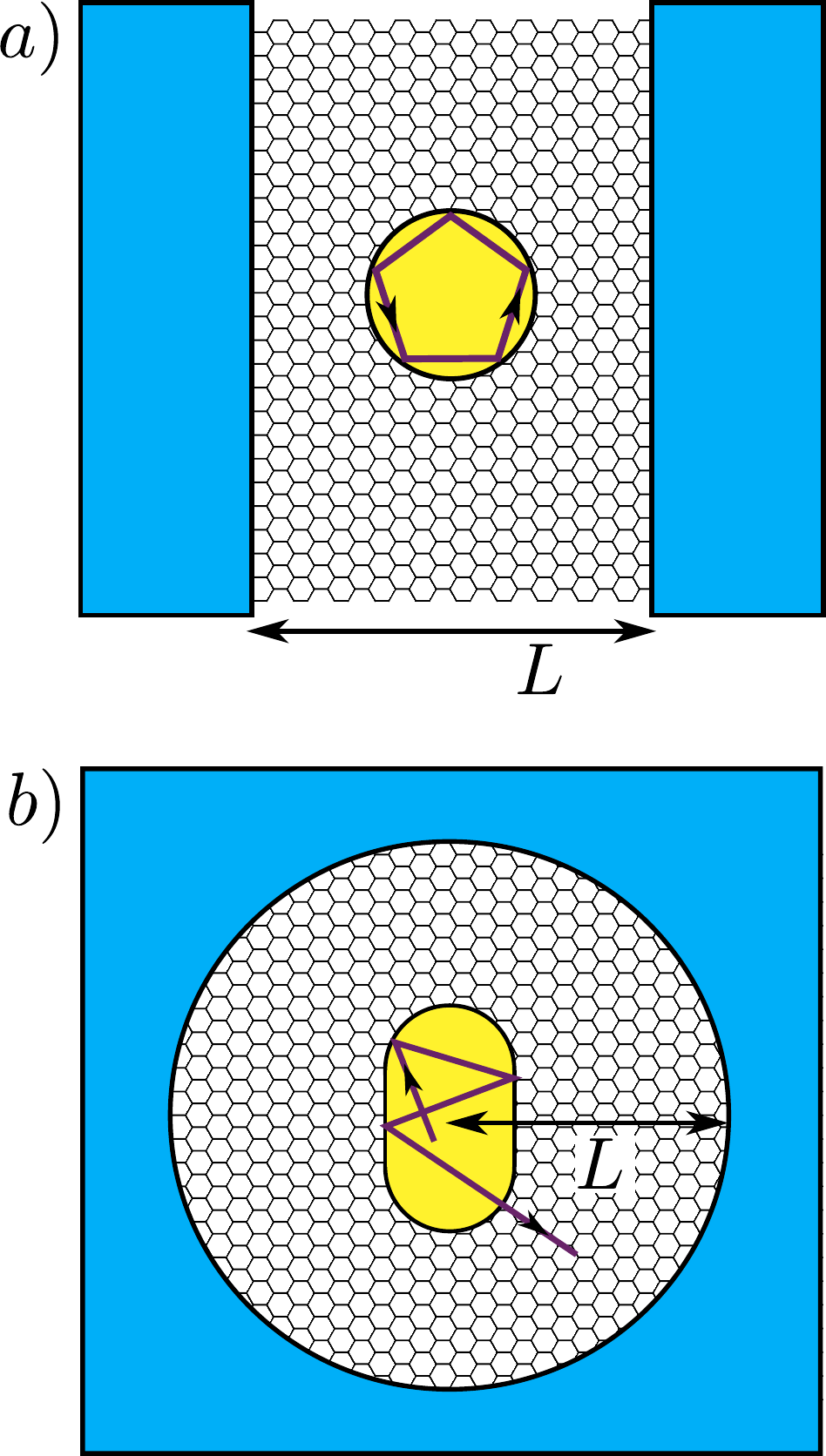}
\caption{(Color online) (a) Gate-defined graphene quantum dot (yellow), surrounded by an intrinsic graphene sheet, and coupled to source and drain reservoirs (blue). (b) The setup considered here: A quantum dot (yellow) surrounded by an undoped graphene sheet, and coupled to a ring-shaped metallic contact (blue). The distance between the dot's center and the contact is denoted $L$. The dot in panel (a) is disc-shaped and has an integrable classical dynamics, for which the reflection angles at the dot's boundary are a constant of the motion. The dot in panel (b) is stadium-shaped. In this case, classical trajectories sooner or later hit the dot's boundary arbitrary close to perpendicular indicence.}
\label{fig:geometry}
\end{figure}

As was pointed out by Bardarson, Titov, and one of the authors,\cite{bardarson2009} this argument leaves open the possibility to electrostatically confine electrons in states in which they are prevented from approaching the boundary of the confinement area at perpendicular incidence.\cite{Egger}
Such a scenario is possible, {\em e.g.}, in a disc-shaped ``quantum dot'' locally gated such that it has a finite carrier density, surrounded by an ``intrinsic'' region of zero carrier density, see Fig.\ \ref{fig:geometry}. Because of the circular symmetry the angle of incidence at the dot's boundary is a constant of the motion, so that electrons in states with nonzero angular momentum are confined inside the quantum dot.\cite{bardarson2009,calvo2011} On the other hand, in geometries with a chaotic classical dynamics one expects that 
there are no bound states, because all classical trajectories eventually approach the boundary arbitrarily close to perpendicular incidence. 

The presence of the pseudospin degree of freedom and the associated Berry phase in graphene calls for a further refinement of this essentially classical argument.\cite{heinl2013} The Berry phase is responsible for a quantization of the kinematic angular momenta $m$ to half-integer values, which excludes the presence of a zero-angular momentum state, that would correspond to strict normal incidence in the Klein tunneling picture. As a result, {\em all} states in a quantum dot, irrespective of its shape, are confined to some degree.\cite{schneider2011,statement}
The insertion of a $\pi$-flux shifts the kinematic angular momentum to integer values, and restores the classical picture of Ref.\ \onlinecite{bardarson2009}, according to which electrostatic confinement is possible in quantum 
dots with integrable classical dynamics only.\cite{heinl2013}

As a quantitative test of confinement, Refs.\ \onlinecite{bardarson2009,pal2011,titov2010,schneider2011,heinl2013} theoretically investigated the two-terminal conductance of an otherwise undoped graphene sheet with a gate-defined quantum dot.\cite{heinisch}
Upon scanning the gate voltage, bound states at zero energy then cause conductance resonances that become narrower if the size $L$ of the undoped graphene sheet containing the quantum dot is increased, see Fig.\ \ref{fig:geometry}. For the disc-shaped quantum dot the resonance widths $\Gamma$ have different $L$ dependencies, $\Gamma \propto 1/L^{2|m|}$, consistent with the expectation that the confinement is stronger for states with higher angular momenta. For a quantum dot with chaotic classical dynamics the widths of all resonances scales with $L$ in the same manner as the broadest resonances of the 
disc-shaped dot, $\Gamma \propto 1/L$, consistent with the semiclassical picture.\cite{schneider2011} Upon insertion of a $\pi$ flux all resonances for the chaotic dot disappear, whereas for the disc-shaped dot only some of the resonances disappear --- the resonances that correspond to the zero-angular-momentum states.\cite{heinl2013}

While the previous studies were focused on the signatures of confinement in the two-terminal conductance,\cite{bardarson2009,schneider2011,heinl2013} the goal of this paper is to extend and complement these studies by investigating how information about confinement is revealed in the density of states or, equivalently, the quantum capacitance of the quantum dot. Density of states measurements provide a valuable experimental technique in the study of nanosystems, and have been also widely used in the context of graphene. One way to gain information is by local probes such as scanning tunneling spectroscopy or scanable single-electron transistors, that give access to the local density of states or the local compressibility of the system, see, {\em e.g.}, Refs.\ \onlinecite{ishigami2007,li2007,martin2008}. Other works also employ capacitive measurements, that extend to the analysis of the total compressibilty of the system, see, {\em e.g.} Refs. \onlinecite{giannazzo2009,xia2009a,droescher2010,ponomarenko2010,
stoller2011}. 
Similar to the study of two-terminal conductance, where the resonant tunneling between dot and leads opens an additional conducting channel and leads to a resonant feature, for the density of states, the presence of well-quantized states in the quantum dot leads to an additional peak structure.

A second motivation to study the density of states, rather than the two-terminal conductance, is of a more technical nature: The theoretical formalism to compute the two-terminal conductance is quite involved,\cite{titov2010,schneider2011} and is not easily adapted to include the $\pi$ flux tube that was used to bring out the role of the Berry phase. For this reason, Ref.\ \onlinecite{heinl2013} performed a direct numerical simulation of the quantum dot structure, which does not allow one to go to the limit of very narrow resonances. On the other hand, measurement of the density of states involves only one contact (in addition to the electrostatic gate) and can be described theoretically without breaking the rotational symmetry. As we show below, this allows for a considerable simplification of the analysis, making it possible to reach the narrow-resonance limit with a $\pi$ flux line, too.

The setup that we study consists of a gate-defined quantum dot surrounded by undoped graphene, and connected to a ring-shaped metallic contact at distance $L$ from the center of the quantum dot, shown schematically in Fig.\ \ref{fig:geometry}b. An unambiguous identification of bound states requires the limit of large $L$, in which the dot is well separated from the metallic contact. It is for this limit that the method presented in this article proves to be particularly effective. Although the rotational symmetry of the leads is chosen primarily for technical reasons, we note that it has no consequence for the qualitative $L$-dependence of the resonances (which can be seen, {\em e.g.}, by comparing the results of the present article with that of Ref.\ \onlinecite{schneider2011}), but also that ring-shaped contacts for graphene can be fabricated in principle,\cite{booth2008} whereas local gating of suspended graphene has also been demonstrated recently.\cite{grushina2013,rickhaus2013}

This paper is arranged as follows: We first consider the signatures of bound states in the density of states for a circular quantum dot in Sec.\ \ref{sec:circqd}. For this situation, a fully analytic solution is amenable. We then contrast the results to the situation of a chaotic quantum dot, which is investigated with the help of a numerical method. In Sec.\ \ref{sec:flux} we complement our studies by introducing a $\pi$-flux into the system, in order to understand the effect of the Berry phase. In this case, we are able to considerably go beyond our previous studies, by going deep to the regime of quantum dots that are weakly coupled to the leads. We conclude in Sec. \ref{sec:conclusion}.

\section{Graphene quantum dot}
\label{sec:circqd}

The setup that we study consists of a gate-defined graphene quantum dot, surrounded by an intrinsic graphene layer, which is connected to a ring-shaped metallic contact. The electronic wavefunction satisfies the two-dimensional Dirac equation
\begin{equation}
 \label{eq:Dirac}
  H \psi_{\eps} = \eps \psi_{\eps}, \ \
  H = v_{\rm F} \vp \cdot \vsigma +V(\vr),
\end{equation}
where $\vF$ is the Fermi velocity, $\vsigma=(\sigma_x,\sigma_y)$ are Pauli matrices, $\vp\equiv-i\hbar\vnabla$ is the momentum operator, and $\eps$ is the energy of the quasiparticle. We take the gate potential to be of the form
\begin{equation}
 V(\vr)=\begin{cases}
         -\hbar v_{\rm F} V_0, & \vr\in {\cal R}\\         
         - \hbar v_{\rm F} V_{\infty}, &r>L, \\
         0,  & \mbox{else}, \\
        \end{cases}
\end{equation}
where, for definiteness, we choose the parameters $V_0$ and $V_{\infty}$ to be positive, such that dot and lead region are electron-doped. The region ${\cal R}$ denotes the area of the dot. For a circular dot ${\cal R}$ consists of all coordinates $\vr$ with $r < R$, $R$ being the dot radius. The ring-shaped metallic contact for $r > L$ is modelled by taking the limit $V_{\infty} \to \infty$. While our choice of a piecewise uniform potential considerably simplifies the calculations, it is not necessary for the existence of bound states,\cite{downing2011,mkhitaryan2012} and our general conclusions will remain valid in the more general case of a central potential $V(r)$. 

Scattering states can be defined in the ring-shaped ideal contact, see Eq.\ (\ref{eq:psiref}) below. In order to calculate the density of states, we use the relation between the local density of states $\nu(\vr,\eps)$ and the derivative of the scattering matrix $\mathcal{S}(\varepsilon)$ with respect to the potential $V(\vr)$ at position $\vr$,\cite{langer1961,buettiker1993b,buettiker1994}
\begin{equation}
  \nu(\vr,\eps) = -\frac{1}{2 \pi i} \mbox{tr}\, {\cal S}^{\dagger} \frac{\delta {\cal S}}{\delta V(\vr)}.
\end{equation}
The total density of states of the dot is then obtained by integration over the region $r < L$,\cite{footdos}
\begin{equation}
  \nu_{\rm dot}(\eps) = -\frac{1}{2 \pi i} 
  \int_{r < L} d\vr \mbox{tr}\, {\cal S}^{\dagger} \frac{\delta {\cal S}}{\delta V(\vr)}.
  \label{eq:nuWS}
\end{equation}
The density of states $\nu_{\rm dot}$ is defined as the density of states per spin and valley.
The expression (\ref{eq:nuWS}) is related to the Wigner-Smith time delay.\cite{wigner1955,smith1960} It is also related to the dot's capacitance,\cite{buettiker2000} which can be measured from the current response to an alternating bias on the ring-shaped metallic contact, at fixed value of the gate voltage $V_0$.

The calculation of the density of states $\nu_{\rm dot}$ at zero energy as a function of the dot potential $V_0$ requires to solve the Dirac equation \eqref{eq:Dirac} at small, but finite energy $\eps$. Dealing with a finite energy also in the region between dot and lead goes beyond previous studies that adressed the two-terminal conductance.\cite{bardarson2009,titov2010,schneider2011,heinl2013}

\subsection{Circular quantum dot}
\label{sec:2A}

We begin our discussion with a circular quantum dot, where the region ${\cal R}$ equals a disc of radius $R$ centered at the origin. In this case the angular momentum $j_z =(\vr\times\vp)_z+\tfrac{\hbar}{2}\sigma_z$ is conserved, and the solutions of the Dirac equation (\ref{eq:Dirac}) can be labeled by the angular momentum $m \hbar$, where $m$ is half-integer as a consequence of the pseudospin degree of freedom for graphene. Writing Hamiltonian (\ref{eq:Dirac}) in polar coordinates,
\begin{equation}
  \label{eq:Hpolar}
  H = -i \hbar v_{\rm F}\matr{0}{\partial_-}{\partial_+}{0} + V(r),
\end{equation}
with the operators
\begin{equation}
 \partial_{\pm}=e^{\pm i \theta} \left(\partial_r \pm i \tfrac{1}{r} \partial_{\theta} \right),
\end{equation}
we solve the Dirac equation $H \psi_{\eps} = \eps \psi_{\eps}$ in the three regions $0 < r < R$, $R < r < L$ and $r > L$ in which the potential $V$ is constant. In each region, we obtain two linearly independent solutions,
\begin{eqnarray}
  \psi_{k,m}^{(\pm)}(\vr) &=& e^{i m \theta} \sqrt{\frac{k}{8 \vF}}
  \nonumber \\ && \mbox{} \times \vect{e^{-i\theta/2 }H^{(\pm)}_{|m-1/2|}(k r)}{i\,\mathrm{sgn}(m)e^{i\theta/2}H^{(\pm)}_{|m+1/2|}(k r)},
 \label{eq:psiref}
\end{eqnarray}
which describe incoming ($-$) or outgoing ($+$) circular waves of wavenumber $k = (\eps - V)/\hbar \vF$ in the conduction band, normalized to unit flux. (Without loss of generality we assume that the energy $\eps$ is positive; We checked that our final results remain valid for negative $\eps$.) Further, the $H_n^{(\pm)}$ are Hankel functions of the first ($+$) and second kind ($-$), respectively. The Hankel functions are related to the Bessel (Neumann) function $J_n$ ($Y_n$) as $H_n^{(\pm)}=J_n\pm iY_n$. 

The precise value of the wavenumber $k$ is different for the three regions in which the solutions (\ref{eq:psiref}) apply. For $r < R$ one has $k \equiv k_0 = \eps/\hbar \vF + V_0$; for $R < r < L$ one has $k=\eps/\hbar \vF$, and for $r > L$ one has $k \equiv k_{\infty} = \varepsilon/\hbar \vF + V_{\infty}$. For $r > L$ the wavefunction can be written as a linear combination of the two solutions of Eq.\ (\ref{eq:psiref}),
\begin{equation}
 \label{eq:psi}
  \psi_{\eps,m}(\vr)= a_{m}(\eps) \psi^{(-)}_{k_{\infty},m} (\vr)+b_{m}(\eps) \psi^{(+)}_{k_{\infty},m} (\vr).
\end{equation}
The coefficients $a_{m}(\eps)$ and $b_{m}(\eps)$ can be determined using continuity of the wavefunction at $r=L$ and $r=R$, as well as regularity at $r=0$. They define the scattering matrix $\mathcal{S}_{mn}(\eps) = \mathcal{S}_{m}(\eps) \delta_{m,n}$ through the relation
\begin{equation}
  \label{eq:scatt}
  b_{m}(\eps) = \mathcal{S}_{m}(\eps) a_{m}(\eps).
\end{equation}
The scattering matrix $\mathcal{S}$ is then used to calculate the density of states, see Eq.\ (\ref{eq:nuWS}).

To simplify the further analysis, we consider the limit of a highly doped lead $k_{\infty} L\gg 1$. In this regime, we make use the asymptotic behavior of the Hankel functions for large arguments, $H^{(\pm)}_n(x)\approx (2/\pi x)^{1/2} e^{\pm i(x-n\frac{\pi}{2}-\frac{\pi}{4})}$ for the wavefunction in the lead region $r > L$. The smallness of the energy $\eps$ furthermore allows to expand in the wavenumber $k$ in the region $R < r < L$ corresponding to the undoped layer separating the quantum dot from the lead. One then finds
\begin{align}
  \mathcal{S}_{m}(\eps) = e^{-2 i k_{\infty} L + i |m| \pi}
  \left[ \mathcal{S}_{m}^{(0)} + k L \mathcal{S}_{m}^{(1)}
  + {\cal O}(\eps^2) \right],
\end{align}
where $k$ is the wavenumber in the region $R < r < L$,
\begin{align}
 \label{eq:calS}
  \mathcal{S}^{(0)}_{m}=&\frac{L^{2|m|}+i {\cal J}_m R^{2|m|}}
  {L^{2|m|} -i {\cal J}_m R^{2|m|}},
\end{align}
and
\begin{widetext}
\begin{align}
  \label{eq:calS1m}
  \mathcal{S}^{(1)}_{m}=&
  \displaystyle
  -\frac{2 i}{2|m|-1} {\cal S}^{(0)}_m
  +
  \frac{8 i |m| L^{4 |m|} +2 i[(2|m|+1){\cal J}_m^2-(2|m|-1)]R^{2|m|+1} L^{2|m|-1}}{(4 |m|^2 - 1)(L^{2 |m|} - i {\cal J}_m R^{2|m|})^2}
\end{align}
\end{widetext}
if $|m| \neq 1/2$, whereas
\begin{align}
  \mathcal{S}^{(1)}_{\pm 1/2}=&
  \frac{i(L^2-R^2)+2 i{\cal J}_{\frac{1}{2}}^2 R^2 \log (L/R)}{(L-i{\cal J}_{\frac{1}{2}}R)^2}.
 \label{eq:S1m12}
\end{align}
In Eqs.\ (\ref{eq:calS})--(\ref{eq:S1m12}) we used the abbreviation
\begin{equation}
 {\cal J}_m=\frac{J_{|m|+1/2}(k_0 R)}{J_{|m|-1/2}(k_0 R)}.
\end{equation}

We now use Eq.\ \eqref{eq:nuWS} to calculate the density of states $\nu_{\rm dot}$ at zero energy,
\begin{align}
 \label{eq:deltanu}
  \nu_{\rm dot} = \frac{1}{2 \pi i \hbar \vF} \sum_{m}
  \mathcal{S}^{(0)*}_m \left[ \frac{\partial \mathcal{S}^{(0)}_m}{\partial V_0}
  + L \mathcal{S}^{(1)}_m \right].
%
\end{align}
The first term in Eq.\ (\ref{eq:deltanu}) represents the integral of the local density of states inside the quantum dot region $r < R$; the second term is the integral of the local density of states in the undoped layer that separates the dot and the metallic contact.


Let us now analyse the density of states as a function of the gate voltage $V_0$. In the limit $R \ll L$ (weak coupling to the ring-shaped contact), the DOS exhibits isolated resonances at gate voltages $V_0 = V_0'$ satisfying the condition 
\begin{equation}
 \label{eq:resonanceposition}
 J_{|m|-1/2}(V_0' R)=0.
\end{equation} 
Close to resonance, we have that ${\cal J}_{m}\approx -1/[R(V_0 - V_0')]$, and the density of states has a Lorentzian dependence on $V_0$. For a generic resonance with $|m| \neq 1/2$ the zero-energy density of states has the form
\begin{align}
 \label{eq:deltanures}
  \nu_{\rm dot} = \frac{4 R |m|}{\pi \hbar \vF (2|m|-1)}
  \frac{\Gamma}{4 R^2 (V_0 - V_0')^2 + \Gamma^2}, 
\end{align}
where the dimensionless resonance width is given by
\begin{equation}
  \label{eq:Gamma}
  \Gamma=2 (R/L)^{2|m|}. 
\end{equation}
%
Resonances are well separated if $R \ll L$. In the lowest angular momentum channel $|m| = 1/2$ the expression for the density of states reads
\begin{align}
  \nu_{\rm dot} = \frac{2 R}{\pi \hbar \vF}
  \left(1 + \log \frac{L}{R} \right)
  \frac{\Gamma}{4 R^2 (V_0 - V_0')^2 + \Gamma^2}, 
\end{align}
with $\Gamma = 2 R/L$.
We remark that the position of the resonances, as well as the scaling of the width agree with the results for the two-terminal conductance,\cite{bardarson2009,titov2010} where $L$ is the distance between source and drain.

We note, that the resonant part of $\nu_{\rm dot}$ that comes from the first term in Eq.\ (\ref{eq:deltanu}) integrates to $1/\hbar \vF$, when integrated over $V_0$,
corresponding to a $2 \pi$ shift of the scattering phase upon tuning $V_0$ through a resonance. The second term in Eq.\ (\ref{eq:deltanu}) gives an additional contribution to the density of states, whose weight decays for higher angular momentum. Remarkably, in the lowest angular momentum channel, this additional contribution has a large factor $\log(L/R)$ factor in the prefactor. We relate the presence of this large factor to the fact, that the bound states of the lowest angular momentum have only a slow decay $\propto 1/r$, such that the wavefunction is marginally non-normalizable.\cite{bardarson2009}

\begin{figure}[t]
\includegraphics[width=2.9in]{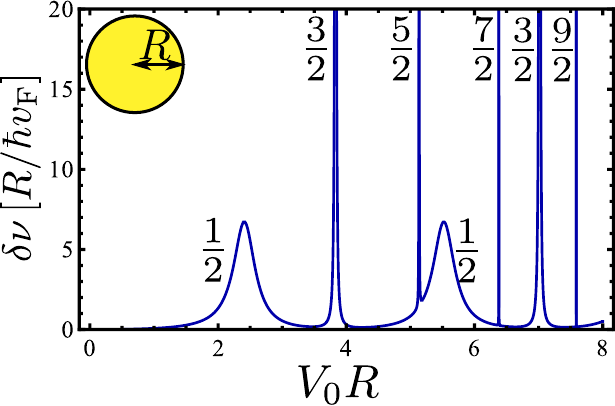}
\caption{(Color online) Density of states for a circular quantum dot. Resonances are labelled according to their angular momentum $|m|$ ($R/L=0.2$).}
\label{fig:circle}
\end{figure}

We show a plot of the density of states for a circular quantum dot as a function of the gate voltage in Fig.\ \ref{fig:circle}. As discussed, the DOS exhibits resonant peaks, that can be labelled according to their angular momentum channel, and the position is given by Eq.\ \ref{eq:resonanceposition}. The higher the angular momentum, the sharper the resonances, consistent with the scaling of the width Eq.\ \eqref{eq:Gamma}.

\subsection{Chaotic quantum dot}
\label{sec:stadium}

We now extend our analysis to a quantum dot of arbitrary geometry. For the generic situation, the scattering matrix is no longer diagonal in the angular momentum basis and an analytical solution is no longer available. We pursue a numerical approach for the calculation of the scattering matrix instead. 

\begin{figure}[t]
\includegraphics[width=2.9in]{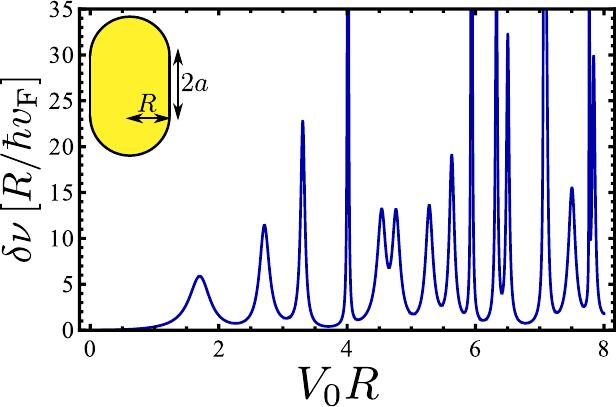}
\caption{(Color online) Density of states for a stadium quantum dot. ($R/L=0.2,2a/R=\sqrt{3}$.)}
\label{fig:stad}
\end{figure}

Our numerical method follows the calculation of the scattering matrix in Ref.\ \onlinecite{schneider2011}. The problem is broken up into thin circular slices, for which the scattering effect is weak and may be captured in Born approximation. The scattering matrix of the full system is then obtained by subsequent concatenation of the scattering matrices of the slices. A difference with Ref.\ \onlinecite{schneider2011} is that we have to calculate the scattering matrix at a finite energy or potential in order to evaluate Eq.\ \eqref{eq:nuWS}. Since the quantum dot has a finite size, the numerical evaluation is necessary up to a distance $\tilde{R}$ away from the origin only. (For the geometry shown in Fig.\ \ref{fig:stad}, one has $\tilde{R}=R+a$). For $\tilde{R} < r < L$ the analytical calculations outlined above can be used. We refer to the appendix for further details of the numerical implementation.

As a prototypical example of a chaotic dot, we now investigate the density of states for a stadium-shaped quantum dot. In Fig.\ \ref{fig:stad}, we show the result of a calculation of the density of states as a function of the gate voltage $V_0$. We find a series of resonances of similar width. The width is comparable to that of the broadest resonances for the circular dot, in agreement with the general expectation that confinement is suppressed in chaotic dots.

\begin{figure}[t]
\includegraphics[width=2.9in]{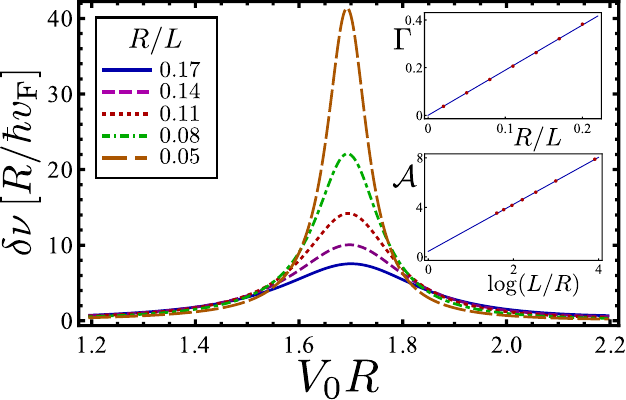}
\caption{(Color online) Density of states for the first resonance, as well as resonance height ${\cal A}$ and width $\Gamma$ (insets) of the stadium dot for various values of the ratio $R/L$. ($2a/R=\sqrt{3}$.)}
\label{fig:firstreso}
\end{figure}

To further analyze the situation, we zoom in on the first resonance and investigate its dependence on the strength of the dot-lead coupling $R/L$, see Fig.\ \ref{fig:firstreso}. We extract height and width of the resonance by fitting to a Lorentzian,
\begin{equation}
 \label{eq:lorentzian}
  \nu_{\rm dot} = \frac{2 R {\cal A}}{\hbar \pi \vF} \frac{\Gamma}{4 R^2 (V_0-V_0')^2+\Gamma^2},
\end{equation}
where $V_0'$ is the resonance position. For the chaotic structure, we expect the resonant states to be composed as a mixture of all angular momentum channels. In the limit of large $L/R$ we expect that the lowest possible angular momentum channel $|m|=\frac{1}{2}$ is dominant. The behavior of the resonances should then resemble those of a $|m|=\frac{1}{2}$--resonance of a circular dot, {\em i.e.}, we expect width and amplitude to scale as
\begin{align}
 \label{eq:GammaA}
 \Gamma&= a\frac{R}{L},\\
   {\cal A}&=b+c\log\frac{L}{R},
\end{align}
with coefficients $a$, $b$, and $c$ of order unity. The numerical study indeed verifies this assertion, as can be seen from the inset of Fig.\ \ref{fig:firstreso}. We further checked, that the other resonances of Fig.\ \ref{fig:stad} show the same behavior for sufficiently small values of $R/L$, although the onset of the asymptotic small-$R/L$ behavior and the precise values of the numerical coefficients $a$, $b$, and $c$ vary from resonance to resonance. We attribute these variations to the different constitutions of the resonances, indicating the relative weigth of the $|m|=\frac{1}{2}$ channel for a certain resonance in comparison to higher angular-momentum channels. We further verified, that the position of the resonances agrees with the ones obtained in a calculation of the conductance (Ref.\ \onlinecite{schneider2011}).

To summarize: For a regular quantum dot, we find signatures of well-confined states, that become very sharp in the limit of weak coupling between dot and lead, as well as broad resonances with a width scaling $\propto R/L$ upon changing the coupling to the lead. For the chaotic dot, we observe such ``broad'' resonances only. The results for the density of states are consistent with the results for a two-terminal conductance setup.

\section{Effect of a $\pi$-flux}
\label{sec:flux}

The difference between regular and chaotic quantum dots in graphene becomes much more pronounced, when we introduce a magnetic flux carrying half a flux quantum. Electrons encircling this flux tube acquire an Aharonov-Bohm phase of $\pi$, which cancels the Berry phase that the electronic wavefunction collects via the pseudospin upon performing a circular motion. The magnetic flux shifts the kinematic angular momentum to integer values, allowing for a state that cannot be confined by gate potentials.

We now analyse the density of states for a graphene quantum dot in the presence of such $\pi$-flux. For this, we add the vector potential
\begin{equation}
\label{eq:flux}
  \vA(\vr)=\frac{\hbar}{e} \frac{1}{2 r} \hat{\vec{e}}_{\theta}.
\end{equation}
Inclusion of $\vA(\vr)$ into the Dirac equation amounts to making the replacement $\partial_{\pm} \to D_{\pm}$ in Eq.\ (\ref{eq:Hpolar}), with
\begin{equation}
 D_{\pm}=e^{\pm i \theta} \left(\partial_r \pm i \tfrac{1}{r} \partial_{\theta} \mp \tfrac{1}{2r} \right).
\end{equation}
We further introduce the kinematic angular momentum,
\begin{equation}
  j_{z,{\rm kin}}=[\vr\times(\vp+e\vA)]_z+\tfrac{\hbar}{2}\sigma_z
\end{equation}
For the $\pi$-flux \eqref{eq:flux}, the kinematic angular momentum is related to the canonical angular momentum as $j_{z,\rm{kin}}=j_z+{\hbar}/{2}$, and therefore is quantized in integer multiples of $\hbar$. We now consider the effect of a $\pi$-flux line on the density of states in a circular and chaotic quantum dot separately.

\subsection{Circular dot}

For the circular dot the flux line is positioned in the origin, so that rotational symmetry is preserved.
The calculation for the density of states proceeds in an analogous way as in the case without flux line. The presence of the flux modifies the basis wavefunctions \eqref{eq:psiref}. We label the new basis states by the integer index of the kinematic angular momentum $\mu$,
\begin{equation}
  j_{z,{\rm kin}} \psi^{(\pm)}_{k,\mu}=\mu\hbar \psi^{(\pm)}_{k,\mu}.
\end{equation}
For non-zero $\mu$, the basis states now read
\begin{equation}
 \label{eq:psirefmu}
 \psi_{k,\mu}^{(\pm)}(\vr)=\sqrt{\frac{k}{8 \vF}} \vect{e^{i(\mu-1)\theta }H^{(\pm)}_{|\mu-1/2|}(k r)}{i\,\mathrm{sgn}(\mu)e^{i\mu\theta }H^{(\pm)}_{|\mu+1/2|}(k r)},
\end{equation}
while for zero kinematic angular momentum, we find
\begin{align}
 \label{eq:psirefmu0}
 \psi_{k,0}^{(\pm)}(\vr)
   =\frac{e^{\pm ikr}}{\sqrt{4 \pi r \vF}} \vect{\pm e^{-i\theta}}{1}.
\end{align}
The state with zero kinematic angular momentum needs to be discussed separately, and will be responsible for the crucial difference caused by the magnetic flux.
Let us discuss the states with non-zero kinematic angular momentum first, where the magnetic flux only leads to slight modifications. Indeed, one finds that the results of Eqs.\  \eqref{eq:calS}-\eqref{eq:calS1m} remain valid, provided the half-integer index $m$ is replaced by the integer index $\mu$, which labels kinematic angular momentum. In particular, resonances in the density of states now appear at roots of half-integer Bessel function $J_{|\mu|-1/2}(V_0 R)=0$, and the resonance width is $\Gamma=2(R/L)^{2|\mu|}$.

For the case $\mu=0$ regularity of the wavefunction at the origin is not sufficient to determine the scattering matrix $\mathcal{S}_0(\eps)$. This problem can be cured by a suitable regularization of the flux line. Taking a flux line of extended diameter we find the condition that the upper component of the wavefunction has to vanish at the origin.\cite{heinl2013} With this regularization the calculation of the scattering matrix $\mathcal{S}_0(\eps)$ is straightforward and has the result
\begin{align}
\mathcal{S}_{0}=e^{-2i(k_{\infty}-k_0)R}e^{-2i(k_{\infty}-k)(L-R)},
 \end{align}
where $k= \eps/\hbar \vF$ and $k_0 = \eps/\hbar \vF + V_0$.
This scattering matrix gives a constant, non-resonant contribution to the density of states, which will be disregarded in the considerations that follow because it is independent of the gate voltage $V_0$.

\begin{figure}[t]
\includegraphics[width=2.9in]{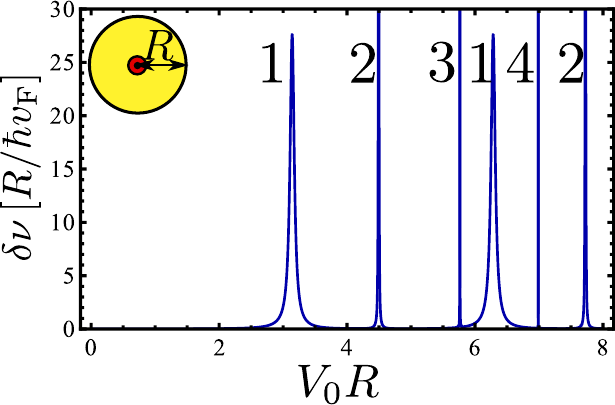}
\caption{(Color online) Density of states for a circular quantum dot in the presence of a flux tube. Resonances are labelled according to their kinematic angular momentum $|\mu|$. ($R/L=0.2$.)}
\label{fig:circleflux}
\end{figure}

We show the density of states for a circular quantum dot in the presence of a flux tube in Fig.\ \ref{fig:circleflux}. It contains resonances originating from non-zero angular momentum channels. The position and the width of such resonances has been discussed above. The zero-angular momentum channel has no $V_0$-dependent contribution to the density of states. Our findings are consistent with a simulation of the two-terminal transport, concerning the position and the scaling of the width upon changing $R/L$ of the resonances,\cite{heinl2013} although the regime of small $R/L$ could not be accessed there.

\subsection{Chaotic Dot}

The numerical method described in Sec.\ \ref{sec:stadium} and the appendix can be easily carried over to the case with the $\pi$-flux, by taking the wave functions Eqs.\ \eqref{eq:psirefmu} and \eqref{eq:psirefmu0} instead of Eq.\ \eqref{eq:psiref}. One has to pay attention to the boundary condition at the origin for the zero angular momentum channel, as discussed above.

\begin{figure}[t]
\includegraphics[width=2.9in]{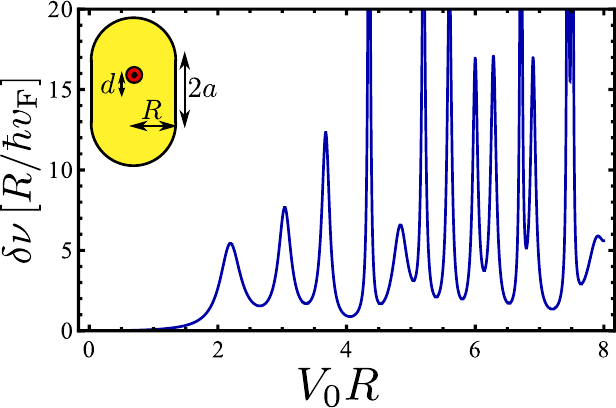}
\caption{(Color online) Density of states for a stadium-shaped quantum dot in the presence of a $\pi$-flux tube. The flux is shifted from the center of the stadium in order to break inversion symmetry und obtain a truly chaotic structure, see also Ref.\ \onlinecite{heinl2013}. ($R/L=0.2,2a/R=\sqrt{3},d=2a/3$.)}
\label{fig:stadflux}
\end{figure}

\begin{figure}[t]
\includegraphics[width=2.9in]{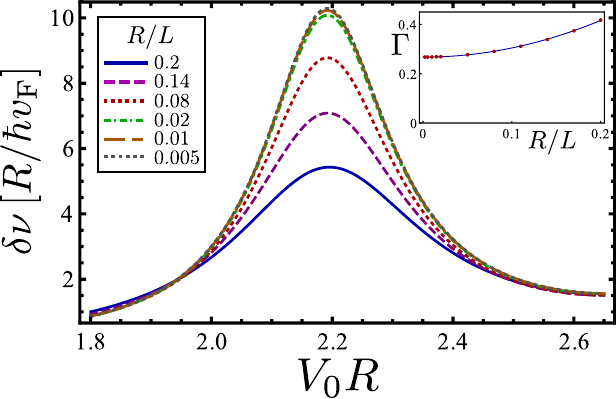}
\caption{(Color online) Density of states for the first ``resonance'' and the corresponding resonance width (inset) of the stadium dot for various values of the ratio $R/L$. Both height and width of the feature saturate in the limit $R/L \to 0$. ($2a/R=\sqrt{3},d=2a/3$.)}
\label{fig:firstresoflux}
\end{figure}

We show the result of a calculation of the density of states of a stadium-shaped quantum dot in the presence of a flux tube in Fig.\ \ref{fig:stadflux}, as a function of the dot's potential. The flux tube is placed off-center in order to lift a twofold rotation symmetry. We observe a density of states with broad peaks, the peak widths typically being much larger than for the circular dot (compare with Fig.\ \ref{fig:circleflux}). The analysis can be made quantitative by considering a specific ``peak'' as a function of $R/L$, see Fig.\ \ref{fig:firstresoflux}. Remarkably, the density of states saturates in the limit $R/L \to 0$ that corresponds to a weak coupling between the quantum dot and the ring-shaped metallic contact. This behavior is a qualitative difference with the case of a circular dot and clearly distinguishes resonances of a chaotic dot with flux from those of a regular dot or the situation without flux. We explain this feature by the special role of the zero angular momentum channel, that 
becomes dominant decay channel in the limit of small $R/L$. As this channel is not capable of binding (or backscattering) states, the density of states becomes insensitive of the distance $L$ to the metallic contact. The transient behavior for $R/L$ of order unity is attributed to the contribution from finite-angular-momentum channels to the resonance width, which gradually disappears if $R/L$ becomes small.

For the specific resonance shown in Fig. \ref{fig:firstresoflux}, the saturation of the density of states takes place only for very small dot sizes. On the other hand, the absence of a bound state may also be inferred from an analysis of the width of the resonance-like feature as a function of $R/L$. Obtaining the width $\Gamma$ from a Lorentzian fit, the inset of Fig.\ \ref{fig:firstresoflux} shows that $\Gamma$ approaches a finite value as $R/L$ goes to zero, with a leading correction $\propto (R/L)^2$, which we attribute to a contribution from the angular momentum channel $|\mu|=1$. Such behaviour is in stark contrast to the scenario without flux tube, where the width goes to zero in the limit $R/L \to 0$. We verified the same qualitative behavior for the other resonance-like features shown in Fig.\ \ref{fig:stadflux}, where the value of $R/L$, at which the width saturates, varies considerably for different resonances, expressing the variations of the relative contribution in the channel of zero angular 
momentum. We also found rough agreement in the positions of the resonances with the ones obtained in the calculation of the conductance (Ref.\ \onlinecite{heinl2013}), although no precise comparison is possible here, since the resonances cannot be made arbitrarily narrow upon making the dot smaller. We note, that our calculation of the density of states allows us to access for much smaller values of $R/L$ as compared to the numerical study of the two-terminal conductance in Ref.\ \onlinecite{heinl2013}, including access to the regime, where the lineshape of the density of states saturates.

\section{Conclusion}
\label{sec:conclusion}

Although graphene has no gap in its electronic excitation spectrum, it is possible to confine electrons electrostatically in a gated region of finite carrier density, if that region is surrounded by a large undoped graphene sheet and if its shape or potential are such, that the electron dynamics is integrable. In order to make this statement quantitative, previous publications investigated the conductance of such a quantum dot, if placed between two metallic electrodes.\cite{bardarson2009,titov2010,schneider2011} In this article we explore the density of states as an alternative signature of confinement. As we have shown, the calculation of the density of states is significantly easier than the calculation of the two terminal conductance, which allows us to extend the analysis to quantum dots with a flux line (in order to highlight the role of the Berry phase associated with the graphene pseudospin) and to access the regime of well isolated resonances, which requires the limit that the metallic contacts are 
far away from the quantum dot. This limit could not be reached in numerical simulations of the two-terminal transport of Refs.\  \onlinecite{bardarson2009,heinl2013}.

The geometry strongly influences the capability of the quantum dot to confine states. While both regular and chaotic dot have resonant signatures in the density of states, the scaling of the width of these features with the (linear) size $L$ of the undoped graphene layer separating the dot and the metallic contacts allows to discern the geometries: While for the chaotic dot, all resonances have a width scaling proprotional to $1/L$, indicating a weak confinement, for the regular dot, most of the resonances have a width that vanishes faster upon increasing $L$, indicating well-confined states.\cite{schneider2011} The difference between chaotic and integrable geometries becomes much more pronounced, when a flux tube carrying half a flux quantum is introduced to the system. In this case, in the limit of large $L$ the lineshape for the density of states for the chaotic dot saturates --- corresponding to a ``resonance width'' that is independent of $L$. Such behavior signals the absence of confined states, in 
contrast 
to the case of a disc-shaped dot, that continues to show sharp resonances after a flux tube has been inserted.

The analysis carried out in this paper considered the density of states integrated over the dot and the surrounding undoped graphene sheet. On the other hand, scanning tunneling microscope experiments measure a local density of states for the region covered by the tunneling tip. The formalism that we developed here can easily be extended to this kind of measurement setup. As far as a qualitative analysis of peak widths and peak positions goes, however, we expect no difference between the local density of states and the integrated density of states that was studied here.

\acknowledgments
We gratefully acknowledge discussions with Silvia Viola Kusminskiy. This work is supported by the Alexander von Humboldt Foundation in the framework of the Alexander von Humboldt Professorship, endowed by the Federal Ministry of Education and Research and by the German Research Foundation (DFG) in the framework of the Priority Program 1459 ``Graphene''.

\appendix
\section{Numerical Method}
\label{sec:numericalmethod}
In this appendix, we provide details concerning the numerical calculation of the scattering matrix for a gate-defined dot or arbitrary shape. We first choose a disc of radius $R$ that fully covers the quantum dot. Inside this disc, {\em i.e.}, for $r < R$, the scattering matrix needs to be determined numerically; outside the disc the Dirac equation can be solved analytically, see Sec.\ \ref{sec:2A}. 

For $r < R$ we rewrite the Dirac equation as
\begin{equation}
 \label{eq:num}
 \left[\vF \vp \cdot \vsigma +U(\vr)\right]\psi=\hbar \vF k_{\rm ref} \psi,
\end{equation}
where $k_{\rm ref}$ is a reference wavenumber that can in principle be chosen arbitrarily and the potential $U(\vr)$ is defined as
\begin{equation}
 \label{eq:ur}
 U(\vr)=V(\vr)-\eps+\hbar\vF k_{\rm ref}.
\end{equation}
We regard Eq.\ \eqref{eq:num} as a scattering problem of an electron with wavenumber $k_{\rm ref}$ on the potential $U(\vr)$. For the solution of this scattering problem, we divide the disc of radius $R$ in $N$ circular slices $r_i<|\vr|<r_{i+1}$, where $0 \equiv r_0 <r_1<r_2< \ldots <r_{N-1}< r_{N} \equiv R$. We first calculate the scattering matrix ${\cal S}^{(i)}$ of the $i$th slice, which is defined for a scattering problem for which the potential $U(\vr)$ is set to zero everywhere except for $r_{i-1} < r < r_{i}$. The scattering matrix ${\cal S}^{(i)}$ is defined with respect to flux-normalized scattering states defined for $r < r_{i-1}$ and $r > r_{i}$, taken at wavenumber $k_{\rm ref}$. If we choose the slices thin enough, a treatment of the scattering problem in the Born approximation is sufficient. The wavenumber $k_{\rm ref}$ should be chosen large enough, that these scattering states are well defined. A too small value of $k_{\rm ref}$ disturbs the numerical procedure, since the Hankel function 
are divergent at small arguments. On the other hand, too large values for $k_{\rm ref}$ require a finer slicing. After the calculation of the scattering matrices ${\cal S}^{(i)}$ for the individual slices, concatenation of those yields the full scattering matrix $\mathcal{S}_{R}(\eps)$ for the potential $U(\vr)$ inside the the disc of radius $R$. This procedure is very similar to the calculation in Ref.\ \onlinecite{schneider2011}, to which we refer for a more detailed description.

For the further calculation, we only need the scattering matrix at small energies. We thus expand
\begin{equation}
 \label{eq:SR0}
 S_{R}(\eps)=S_{R}^{(0)}+S_{R}^{(1)}\eps+O(\eps^2),
\end{equation}
where the matrices $S_{R}^{(0)}$ and $S_{R}^{(1)}$ are obtained from the numerics by setting the energy $\eps$ in the potential $U(\vr)$ (Eq. \eqref{eq:ur}) first to zero, and then to a very small value.

To calculate the density of states, we need the full scattering matrix ${\cal S}$, that relates in- and outgoing states in the metallic lead, {\em i.e.} we still need to account for the undoped graphene region $R < r < L$. Since the problem is circularly symmetric outside the disc of radius $R$, one can establish an explicit connection between ${\cal S}$ and $\mathcal{S}_{R}$. For $r > L$ the wavefunction for an electron incident in angular momentum channel $m$ has the form
\begin{equation}
  \label{eq:psin}
  \psi_{\eps}(\vr) = \psi^{(-)}_{k_{\infty},m}(\vr) +
  \sum_{n} \mathcal{S}_{nm}(\eps) \psi^{(+)}_{k_{\infty},n}(\vr),
\end{equation}
which defines the scattering matrix $\mathcal{S}_{nm}$ in the general case that angular momentum is not conserved.
For $R < r < L$ we may expand the solution of the Dirac equation as
\begin{equation}
   \psi_{\eps}(\vr) = \sum_{n} \left[ a_{nm} \psi^{(-)}_{k,n}(\vr) +
   b_{nm} \psi^{(+)}_{k,n}(\vr) \right],
\end{equation}
with $k = \eps/\hbar v_{\rm F}$, whereas for the limit $r \uparrow R$ the wavefunction may be written as
\begin{equation}
  \psi_{\eps}(\vr) = \sum_{n} c_{nm} \left[ \psi^{(-)}_{k_{\rm ref},n}(\vr)
  + \sum_{p} \mathcal{S}_{R,pn}(\eps) \psi^{(+)}_{k_{\rm ref},p}(\vr) \right],
\end{equation}
with the scattering matrix $\mathcal{S}_{R}(\eps)$ as defined above. By imposing continuity of the wavefunction at $r=R$ and $r=L$, we can eliminate the coefficients $a_{nm}$, $b_{nm}$ and $c_{nm}$ and express $\mathcal{S}(\eps)$ in terms of $\mathcal{S}_R(\eps)$. This program can be carried out analytically, including the expansion in $k$ relevant for the application of Eq.\ (\ref{eq:deltanu}) of the main text. The resulting equations can be obtained in a straightforward manner, but they are too lengthy to be reported here.

We checked that our results do not depend on the choice of $R$ and $k_{\rm ref}$ that were introduced into the numerical procedure as auxiliary parameters.


\begin{thebibliography}{40}
\bibitem{statement} The precise statement is that all states give rise to sharp resonances in a conductance or capacitance measurement, with a width that goes to zero in the limit that the undoped region surrounding the quantum dot becomes large.
\bibitem{footdos} The functional derivative $\delta {\cal S}/\delta V(\vr')$ can be obtained by changing the potential $V(\vr)$ by $v_0 \delta(\vr-\vr')$ and taking the derivative of the scattering matrix with respect to $v_0$. In practice, for the calculation of the density of states integrated over a certain region, one shifts the potential $V(\vr)$ by a constant $v_0$ within the specific region, and takes the derivative of the scattering matrix with respect to $v_0$.
\end{thebibliography}
\end{document}